\documentclass[aps,groupedaddres8s,fleqn,nofootinbib,twocolumn]{revtex4}
\usepackage{graphicx}
\usepackage{xcolor}
\usepackage{amsmath,amssymb}
\usepackage{float}
\usepackage{amsfonts}
\usepackage{verbatim}
\usepackage{flushend}
\usepackage{balance}
\usepackage{endnotes}
\usepackage{footnote}
\usepackage{adjustbox}
\usepackage{gensymb}
\usepackage{float}
\usepackage{epigraph}
\usepackage{xcolor}
\usepackage[utf8]{inputenc}
\usepackage{setspace}

\begin{document}

\author{K. Trachenko}
\thanks{k.trachenko@qmul.ac.uk}
\affiliation{School of Physical and Chemical Sciences, Queen Mary University of London, 327 Mile End Road, London, E1 4NS, United Kingdom}

\title{Fundamental physical constants, operation of physical phenomena and entropy increase}

\begin{abstract}
Approaching the problem of understanding fundamental physical constants (FPCs) started with discussing the role these constants play in high-energy nuclear physics and astrophysics. Condensed
matter physics was relatively unexplored in this regard. More recently, it was realised that FPCs set lower or upper bounds on key condensed matter properties. Here, we discuss a much wider role
played by FPCs in condensed matter physics: at given environmental conditions, FPCs set the observability and operation of entire physical effects and phenomena. We discuss structural and
superconducting phase transitions and transitions between different states of matter, with implications for life processes. We also discuss metastable states, transitions between them, chemical
reactions and their products. A byproduct of this discussion is that the order of magnitude of the transition temperature can be calculated from FPCs only. We show that the new states emerging as a
result of various transitions increase the phase space and entropy. Were FPCs to take different values, these transitions would become inoperative at our environmental conditions and the new states
due to these transitions would not emerge. This suggests that the current values of FPCs, by enabling various transitions and reactions which give rise to new states, promote entropy increase.
Based on this entropy increase and the associated increase of statistical probability, we conjecture that entropy increase is a selection principle for FPCs considered to be variable in earlier
discussions.
\end{abstract}

\maketitle

\section{Introduction}

Fundamental physical constants (FPCs) determine physical processes at all energy and length scales of our experiments and theories. These different length scales are illustrated in Figure 1. The values of FPCs are often listed in textbooks and reviews (see, e.g., Refs. \cite{codata,ashcroft,uzanreview,uzan1}), and their recommended values are updated and maintained in the National Institute for Standards and Technology database \cite{nist-fund}.

\begin{figure*}
{\scalebox{0.49}{\includegraphics{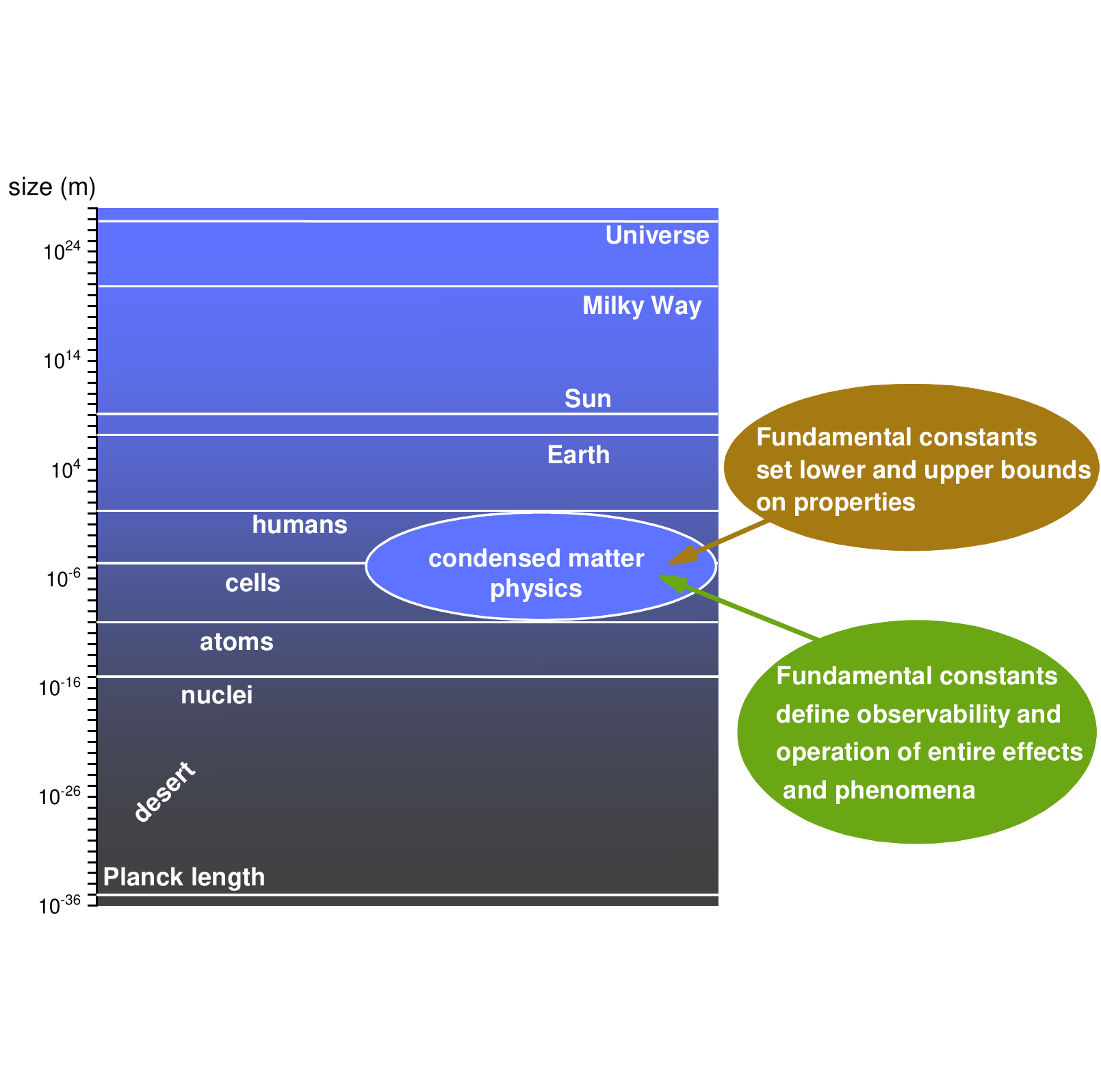}}}
\caption{Different scales of structure in the Universe including the typical size range in condensed matter physics. Fairly recently, FPCs were shown to put lower and upper bounds on different condensed matter properties - properties that are already observed - including viscosity, diffusion and spin dynamics, electron and heat transport, thermal conductivity in insulators and conductors, the speed of sound, elastic moduli, flow in cells, vibration frequency in solids and superconducting temperature (see Introduction). Here, we discuss a wider role of FPCs affecting observability and operation of entire phenomena and effects, including structural and superconducting transitions, transitions between different states of matter and chemical reactions.}
\label{fig1}
\end{figure*}

Together with our fundamental theories, FPCs are the key to understanding the present state of our Universe and its contents as well as its past and future
\cite{barrow,barrow1,carr,carrbook,finebook,cahnreview,hoganreview,adamsreview,uzanreview}. The ``barcodes of ultimate reality'' \cite{barrow}, FPCs also set our Universe apart from those worlds
where these constants might be different.

Understanding the values and origin of FPCs is one of the grandest challenges in modern science \cite{grandest}. The problem of understanding FPCs is not that it involves very hard technical calculations but that we do not know anything more fundamental \cite{weinberg}. We do not know what kind of theoretical tools we need to address the problem. It is probably this enormity of the problem that explains why there seems to be no sustained programme of research in this area, scheduled meetings and so on, as compared to more developed fields of science.

A starting point to understand FPCs is to ask what would happen were FPCs to take values different from those currently observed \cite{barrow,barrow1,carr,carrbook,finebook,cahnreview,hoganreview,adamsreview}? This is done on the basis of our physical models. The bottom in Figure 1 involves length scales of particle physics. According to our theories, observed particles involve a finely-tuned balance between the masses of down and up quarks: larger up-quark mass gives the neutron world without protons and hence no atoms made from nuclei and electrons around them, whereas larger down-quark mass gives the proton world without neutrons where light atoms can form only but not heavy atoms. Our world with heavy atoms and electrons endowing complex chemistry would not exist with a few per cent fractional change in the mass difference of the two quarks \cite{hoganbook,hoganreview}.

The top in Figure 1 involves cosmological scales and a balance between the gravitational and cosmological constants. This balance gives enough time for the Universe to expand and not to contract too quickly and at the same time maintain sufficient gravitational clamping so that galaxies, stars and other structures can form \cite{barrow,adamsreview}.

For the rest of intermediate length scales in Figure 1, our theories imply that there is an intricate network of inter-related constraints on FPCs that need to be satisfied to give many structures in the observable Universe. This includes formation and evolution of different astronomical objects and in particular stars, synthesis of heavy elements in stars including carbon and oxygen and other elements which can then form molecular structures essential to life \cite{barrow,adamsreview}. The strength of these constraints, the degree of tuning between FPCs, depends on how well developed our physical models are. Ref. \cite{adamsreview} discusses existing uncertainties involved in our models and how these uncertainties affect constraints imposed on FPC.


Compared to the above high-energy processes, the discussion of the role of FPCs at the scale of condensed matter has been scarce - this is the scale spanning 10$^8$-10$^9$ orders of magnitude in Figure 1 or about 20\% of the logarithmic scale between the nuclei and the size of Universe. This has started to change fairly recently.

It was realised that FPCs set upper or lower bounds on several key condensed matter properties. Examples include bounds on viscosity \cite{sciadv}, diffusion and spin dynamics \cite{spin,hbarm1,hbarm2,hbarm3,hbarm4}, electron and heat transport \cite{hartnoll,hartnoll1,behnia,nussinov1,grozdanov}, thermal conductivity in insulators \cite{momentumprb} and conductors \cite{brareview}, the speed of sound \cite{sciadv2}, elastic moduli including those in low-dimensional systems \cite{brareview,advphysreview}, flow in cells \cite{sciadv2023,ropp1}, vibration frequency in solids and superconducting temperature \cite{ourtc}. This has deepened our understanding of these properties and their physical models.

Here, we discuss a much wider role of FPCs in condensed matter physics: extending beyond setting bounds on properties that are {\it already observed}, FPCs affect observability and operation of entire physical phenomena and effects. In the next section, we discuss why discussing FPCs at the condensed matter scale is important. This is followed by considering phase transitions including structural and superconducting transitions and transitions between different states of matter with implications for life processes. This shows how FPCs set the operation of entire physical phenomena such as phase and other transitions and hence the existence of different states of matter emerging as a result of these transitions. An interesting byproduct of this discussion is that the order of magnitude of the transition temperature can be calculated from FPCs only.

We show that the new states emerging from various transitions increase the phase space and entropy. We also show that were FPCs to take different values, these transitions would become inoperative at
our environmental conditions, and the new states would not emerge. This suggests that the current values of FPCs, by enabling various transitions and reactions which produce the new states, result in
the increase of the phase space and entropy. We propose that this is a general effect operating at different levels of structure in the Universe. The entropy increase due to FPCs can be discussed in
terms of an evolution-like mechanism where this increase is analogous to selection in biological evolution. Based on this entropy increase and the associated increase of statistical probability, we
conjecture that the entropy increase is a selection principle for FPCs considered to be variable in earlier discussion‌s.


\section{Why filling the condensed matter gap is important}

There are several reasons why understanding the role of FPCs at condensed matter physics scales and filling the condensed matter gap in Figure 1 is important. First and quite generally, having a non-interrupted and continuous set of length and energy scales in Figure 1 over which we can discuss the role of FPCs gives more information and detail to use.

Second, the uncertainty and insufficient development of physical models related to higher-energy processes limits our understanding of FPCs \cite{adamsreview}. On the other hand, condensed matter physics involves a larger variety of ways to test condensed matter theories. Many of these tests are more readily available than those in high energy physics. As a result, theories and concepts in condensed matter physics have been well verified and cross-checked using a great variety of experiments and simulations.

Third, understanding the role of FPCs in condensed matter physics is an important enterprise on its own. It enriches the field itself and deepens understanding condensed matter phenomena at the most basic level. It also provides links and relationships to other areas of physics.

The fourth and final reason discussed here is that if we better understand the operation of FPCs in condensed matter physics, we may find that the underlying principles are quite general and can be applied to other length and energy scales. This can lead to new insights related to the origin of FPCs and their values.

In the next section, we discuss several examples where FPCs set observability and operation of several key condensed matter effects and phenomena.

\section{Observability and operation of phase transitions. Emergence of new states with higher entropy}
\label{observ}

\subsection{Fundamental and environmental parameters}

We start with separating two types of parameters. The first type are FPCs. The second type are the environmental parameters such as temperature $T$ and pressure $P$ at typical planetary conditions where we do our observations. These parameters are not fundamental but are due to historical accidents \cite{weinberg-dreams} such as the distance between the planet and its star which affects the planet's temperature. In the Solar system, for example, $T$ is in the range 10$^2$--10$^3$ K.

In this paper, we will consider how FPCs affect the observability and operation of condensed matter phenomena such as phase transitions at our environmental conditions. However, the discussion and its implications apply to any environmental conditions. To keep the discussion general, below we consider the environmental temperature in the range $T_1$--$T_2$. Ref. \cite{adamsreview} discusses several important astrophysical processes which set the range of environmental conditions.

FPCs discussed in this paper will include those central to condensed matter physics: the Planck constant $\hbar$, electron mass $m_e$ and charge $e$. These constants define the characteristic scales of distance and energy in condensed matter by setting the Bohr radius $a_{\rm B}$ and Rydberg energy $E_{\rm R}$ as

\begin{equation}
a_{\rm B}=\frac{4\pi\epsilon_0\hbar^2}{m_e e^2}
\label{bohr}
\end{equation}

\noindent and

\begin{equation}
E_{\rm R}=\frac{m_ee^4}{32\pi^2\epsilon_0^2\hbar^2}
\label{rydberg}
\end{equation}

$a_{\rm B}\approx 0.5$~\AA\ and is on the shorter side of interatomic separations in condensed phases in the typical range 1--3 \AA. $E_{\rm R}\approx$~13.6 eV and is on the larger side of cohesive energies in condensed phases in the range 1--10 eV.

We will also discuss the mass of the proton (the composite particle), $m_p$, which sets the scale of lattice oscillation frequency (see section \ref{observa}). $m_p$ will mostly enter our discussion as the dimensionless ratio

\begin{equation}
\beta=\frac{m_p}{m_e}
\label{beta}
\end{equation}

We note another important dimensionless parameter, the fine-structure constant

\begin{equation}
\alpha=\frac{1}{4\pi\epsilon_0}\frac{e^2}{\hbar c}
\label{alpha}
\end{equation}

\noindent where $c$ is the speed of light in vacuum. $\alpha$ and $\beta$ play a key role at different length and energy scales in Figure 1, including the stability of particles, nuclear reactions, formation and evolution of stars, synthesis of heavy nuclei and so on. These processes impose constraints on the allowed range of $\alpha$ and $\beta$ (as well as on the strong coupling constant) \cite{barrow,adamsreview}. $\alpha$ will play an indirect role in our discussion in Section \ref{observa} because we don't consider relativistic effects here. This may change if relativistic effects play a role in other condensed matter processes.

This amounts to the following set of FPCs discussed in this paper:

\begin{equation}
{\rm FPCs}=\{\hbar,e,m_e,\frac{m_p}{m_e},\alpha\}
\end{equation}

\subsection{Structural phase transitions}
\label{secondor}

\subsubsection{Entropy increase due to a phase transition}

Our first case study are structural phase transitions in solids. Let's consider common silica SiO$_2$ which has several polymorphs at different pressure and temperature, separated by phase transitions. As a model example, let's consider two silica phases: quartz and high-temperature $\beta$-cristobalite phase \cite{dove-rums1,dove-rums2}.

The solid entropy in the classical high-temperature regime $T\gg\hbar\omega_{\rm D}$, where $\omega_{\rm D}$ is Debye frequency, is \cite{landaustat}:

\begin{equation}
S_s(T\gg\hbar\omega_{\rm D})=3N\left(\ln\frac{T}{\hbar\bar{\omega}}+1\right)
\label{soliden}
\end{equation}

\noindent where $\bar\omega$ is the average phonon frequency set as $\bar\omega^{3N}=\prod\limits_i\omega_i$, $\omega_i$ are $3N$ phonon frequencies, $N$ is the number of atoms and anharmonic corrections are assumed small and not considered below. Here and below, $k_{\rm B}=1$.

As expected for high-temperature phases on general grounds \cite{landaustat}, cristobalite has a more symmetric and open structure compared to quartz. As a result, cristobalite has a significantly larger number of low-frequency phonons, rigid-unit modes, that propagate without the distortion of rigid SiO$_4$ tetrahedra \cite{dove-rums1,dove-rums2}. This gives smaller $\bar\omega$ in cristobalite and larger entropy in Eq. \eqref{soliden}.

The entropy \eqref{soliden} reflects the phase space available to $3N$ phonons in solids, and the increase of entropy of cristobalite is due to the phonon softening. However, the same entropy increase can be seen directly as the increase of the phase space of particle coordinates $q$ and momenta $p$, $\Delta\Gamma$. $\Delta\Gamma$ is:

\begin{equation}
\Delta\Gamma=\frac{1}{(2\pi\hbar)^{3N}}\prod\limits_i\Delta p_i\Delta q_i
\label{gamma}
\end{equation}

Using $\langle\Delta p_i\rangle^2\propto\frac{T}{\omega_i^2}$ and $\langle\Delta q_i\rangle^2\propto\frac{T}{\omega_i^2}$ for each phonon mode in Eq. \eqref{gamma} and

\begin{equation}
S=\ln\Delta\Gamma
\label{log}
\end{equation}

\noindent gives the entropy term $S\propto 3N\ln\frac{T}{\hbar\bar{\omega}}$ as in Eq. \eqref{soliden}.

We see that smaller $\omega_i$ due to softer phonons in cristobalite increase $\langle\Delta p_i\rangle^2$ and $\langle\Delta q_i\rangle^2$. This increases $\Delta\Gamma$ and $S$ of cristobalite compared to quartz.

Frequency softening above the phase transition can be seen more generally in terms of the order parameter $\eta$. Considering that phonon frequencies are renormalised by anharmonicity related to a phase transition, the renormalised frequencies $\tilde\omega_i$ are \cite{dove-rums2,dovesecbook}:

\begin{equation}
\tilde\omega_i^2=\omega_i^2+\frac{1}{2}\delta_i\eta^2
\end{equation}

\noindent where $\omega_i$ are harmonic frequencies and $\delta_i$ are the anharmonicity coefficients.

In the high-temperature phase above the phase transition temperature $T_c$, $\eta=0$. Phonon frequencies become smaller as a result, increasing the entropy.

Importantly, the entropy increase is ultimately related to a phase transition which gives rise to the second high-entropy phase. Without a phase transition operating, the entropy would increase according to Eq. \eqref{soliden} within the single low-temperature phase only but without larger entropy coming from the second high-temperature phase.

We consider $T_c$ to be within the range of environmental temperature, $T_1<T_c<T_2$. The entropy increase due to phase transitions can take place both above and below $T_c$. For reversible phase transition where the low-temperature phase is always recovered below $T_c$, the entropy increase takes place above $T_c$. In our model quartz-cristobalite example, this gives

\begin{equation}
S_C(T>T_c)>S_Q(T>T_c)
\label{ineq1}
\end{equation}

\noindent where $S_C(T>T_c)$ is the entropy of the high-temperature cristobalite phase above $T_c$ and $S_Q(T>T_c)$ is the entropy that quartz would have above $T_c$ without the phase transition operating. More generally, \eqref{ineq1} applies to any two low and high-temperature phases.

The entropy increase can also take place below $T_c$ if the high-temperature phase is a long-lived metastable state below $T_c$. In our quartz-cristobalite example, the phase transition is a reconstructive transition with large energy barriers separating the two phases. As a result, cristobalite is a long-lived metastable phase which can be cooled back to room or any other temperature below $T_c$ \cite{dove-rums1,dove-rums2}. 
We note here that the entropy and other thermodynamic functions of non-equilibrium metastable systems are defined in the commonly used sense: at times shorter than the relaxation time of the non-equilibrium system, this system can be conditionally considered in an equilibrium state where observed properties are described by equilibrium thermodynamics \cite{landaustat}. This is discussed in Section \ref{meta} in more detail.

We find both quartz and cristobalite at our environmental conditions, hence let's consider the entropy of the quartz-cristobalite mixture at any temperature below $T_c$ where both phases co-exist, $S_{Q-C}$. This mixture consists of any proportion of the two phases with the total number of atoms $N$. We compare $S_{Q-C}$ to the entropy of the initial quartz phase, $S_Q$, with the same number of atoms $N$. As discussed earlier, the entropy of cristobalite is higher than that of quartz. Hence,

\begin{equation}
S_{Q-C}(T<T_c)>S_Q(T<T_c)
\label{ineq}
\end{equation}

The increased phase space and entropy above $T_c$ become ``frozen-in'' in the metastable state below $T_c$. As a result, the entropy increase due to phase transitions takes place at all temperatures in the range $T_1$--$T_2$.

The quartz-cristobalite example is readily generalised to many high-temperature polymorphic phases of different solids emerging as a result of phase transitions, with overall entropy increased.

\subsubsection{Observability of phase transitions. Entropy changes.}
\label{observa}

We now observe that if we consider different values of FPCs, the phase transition would not operate. The upper bound on the maximal phonon frequency $\omega_m$ (``cage rattling'' frequency) can be written in terms of FPCs \cite{ourtc}. This bound follows from combining Eq. \eqref{rydberg} and the ratio of $\hbar\omega_m$ and cohesive energy $E$. The ratio involves a harmonic approximation for the relation between $E$ and $\omega_m$ and reads \cite{sciadv,advphysreview}:

\begin{equation}
\frac{\hbar\omega_m}{E}=\left(\frac{m_e}{m}\right)^{\frac{1}{2}}
\label{ratio}
\end{equation}

\noindent where $m$ is the atom mass. Setting $E=E_{\rm R}$ gives

\begin{equation}
\omega_m=\frac{1}{32\pi^2\epsilon_0^2}\frac{m_ee^4}{\hbar^3}\left(\frac{m_e}{m}\right)^{\frac{1}{2}}
\label{f1}
\end{equation}

$\omega_m$ has the upper bound, $\omega_m^u$, for $m=m_p$ as \cite{ourtc}

\begin{equation}
\omega_m^u=\frac{1}{32\pi^2\epsilon_0^2}\frac{m_ee^4}{\hbar^3}\left(\frac{m_e}{m_p}\right)^{\frac{1}{2}} \approx 3680\,\rm {K}
\label{f2}
\end{equation}

$\omega_m^u$ interestingly contains the proton-to-electron mass ratio $\beta$ \eqref{beta}. As mentioned earlier, this ratio is of particular importance for different processes and reactions involving different length scales in Figure 1 \cite{barrow,adamsreview}.

Let's consider what would happen to a phase transition were FPCs to take different values. Let's consider variations of FPCs which increase $\omega_m$ and $\omega_m^u$ in Eqs. \eqref{f1}-\eqref{f2}. Note that this question is well-posed because it is possible to vary $\omega_m^u$ by varying $m_e$ and $m_p$ or $\hbar$ and $e$ while keeping $\beta$ and $\alpha$ in Eqs. \eqref{beta} and \eqref{alpha} unchanged and hence maintain all preceding essential processes (e.g. stellar formation and evolution, nuclear synthesis, stability of ordered structures and so on) intact \cite{barrow,adamsreview}.

Let's consider $\omega_m$ exceeding all temperatures in the considered environmental temperature range $T_1$--$T_2$:

\begin{equation}
T_2\ll\hbar\omega_m
\label{tcond}
\end{equation}

Recalling that $\omega_m$ is close to $\omega_{\rm D}$, condition \eqref{tcond} implies $T\ll\hbar\omega_{\rm D}$. This, in turn, implies that the high-temperature classical result \eqref{soliden} does not apply because the system is now in the quantum regime. In this regime, instead of Eq. \eqref{soliden}, the entropy is \cite{landaustat}

\begin{equation}
S_s(T\ll\hbar\omega_{\rm D})=\frac{4\pi^4}{5}N\left(\frac{T}{\hbar\omega_{\rm D}}\right)^3
\label{entr3}
\end{equation}

$S_s$ is close to 0 when $T\ll\hbar\omega_{\rm D}$. Then the second term in the free energy $F=E-TS$ is close to 0 too, and $F$ now does not have a crossover corresponding to a phase transition. The phase transition becomes inoperative, and there is no associated entropy increase discussed in the previous section.

The observability of phase transitions can also be discussed on the basis of $T_c$. Phase transition theories, including the Ising model and the anharmonic phonon theory (equivalent to Landau theory), expectedly predict that the order of magnitude of $T_c$ is set by the energy of interaction between particles \cite{dove-rums2}. This energy is close to the phonon energy $\hbar\omega_{\rm D}$ or $\hbar\omega_m$. Using Eq. \eqref{f1},

\begin{equation}
T_c\approx\hbar{\omega_m}=\frac{1}{32\pi^2\epsilon_0^2}\frac{m_ee^4}{\hbar^2}\left(\frac{m_e}{m}\right)^{\frac{1}{2}}
\label{tcval}
\end{equation}

To estimate $T_c$, we write $m=Am_p$, where $A$ is the atomic mass. Using Eqs. \eqref{f1} and \eqref{f2}, $T_c$ in Eq. \eqref{tcval} is

\begin{equation}
T_c=\frac{\hbar\omega_m^u}{A^\frac{1}{2}}
\label{tcval1}
\end{equation}

$T_c$ can be estimated by recalling that $A^\frac{1}{2}$ varies across the periodic table in the range of about 1-15, with an average of 8. Using this value and $\omega_m^u=3680$ K \eqref{f2}, Eq. \eqref{tcval1} gives $T_c\approx 500$ K. This is close to $T_c$ derived using typical parameters of the anharmonic phonon model and Landau theory \cite{doveblue} and is of the right order of magnitude of $T_c$ typically observed in the range $10^2$-$10^3$ K.

We see that the order of magnitude of $T_c$ \eqref{tcval} is set by FPCs. Were different values of FPCs to give $\omega_m$ and $T_c$ in Eq. \eqref{tcval}  on the order of, for example, $10^5$ K or higher, phase transitions would be unobservable at our environmental conditions. This is consistent with our earlier result: large $\omega_{\rm D}$ gives small entropy in Eq. \eqref{entr3}, no phase transition and no associated entropy increase.

We therefore find that FPCs, by enabling the operation of phase transitions which give rise to new states, increase the phase space and entropy. We note that phase transitions in the phase-changing subsystem (PCS) involve the heat flow from another, heat-supplying subsystem (HSS) in order to reach $T_c$ and the transition (if the second high-entropy phase in the PCS is metastable below $T_c$, the heat can be reversibly returned from the PCS to HSS below $T_c$, retaining the entropy increase in the metastable phase and involving little or no net heat exchange between PCS and HSS). The PCS and HSS are open, however the combined PCS and HSS subsystem can be considered closed. The entropy increase due to phase transitions in the PCS applies to this closed subsystem and represents the entropy increase which is {\it extra} as compared to the case where no new phases emerge in the PCS and the entropy increase is due to the trivial heat exchange between the PCS and HSS only.

In summary, the discussion in Section \ref{secondor} has so far given us  several insights:

\begin{enumerate}
\item FPCs govern key effects in condensed matter physics and define the regime in which the system is, classical or quantum.
\item FPCs set the observability and operation of entire physical phenomena such as phase transitions (more examples are in Sections \ref{superc}, \ref{firstor} and \ref{meta}).
\item FPCs, by enabling phase transitions, increase the phase space and entropy (more discussion is in Sections \ref{firstor}, \ref{meta} and \ref{entropy} below).
\end{enumerate}

\subsection{Superconducting transition}
\label{superc}

Perhaps unexpectedly at this point, the results in the previous section lead to the question of observability of superconductivity. We have recently asked \cite{ourtc} whether there is a fundamental limit to how high the superconducting temperature, $T_c$, can get? Fundamental limit here specifically means the limit in terms of FPCs.

Using the Migdal-Eliashberg theory, the upper limit to $T_c$, $T_c^u$, can be written as $T_c^u=C\omega_m^u$, where $\omega_m^u$ is given in Eq. \eqref{f2} \cite{ourtc}. $C$ is a function of the electron-phonon coupling constant $\lambda$ and initially increases with $\lambda$ but has a maximum around $\lambda=1-2$ because larger $\lambda$ result in instabilities which lower $T_c$. This range of $\lambda$ corresponds to $C=0.1-0.2$. Using Eq. \eqref{f2}, this gives $T_c^u$ as \cite{ourtc}

\begin{equation}
T_c^{u}=C\omega_m^u=C\left(\frac{m_e}{m_p}\right)^{\frac{1}{2}}\frac{1}{32\pi^2\epsilon_0^2}\frac{m_ee^4}{\hbar^3}\approx 10^2-10^3\,\rm {K}
\label{tc}
\end{equation}

Using Eq. \eqref{tc}, we ask how would the observability of superconductivity change were fundamental constants to take different values? Since $T_c^u$ is the upper bound, $T_c^u$ on the order of 10$^{-5}$ K or lower due to different FPCs implies that superconductivity would be unobserved at our experimental conditions.

Another interesting realisation comes from observing that a variation of FPCs giving $T_c^u$ of, for example, 10$^5$ K or higher, would imply that many materials would have $T_c$ above room temperature. Consistent with Eq. \eqref{tc} giving $T_c^{u}$ on the order of $10^2-10^3$ K due to current FPCs, we observe superconductivity in the range $T\lesssim$ 100-200 K, with the highest $T_c$ seen in hydrides \cite{hreview1,hreview2,hreview3}. This naturally stimulates the current intense research into finding materials with $T_c$ of 300 K and above. We therefore see that the very existence of the current research to find systems with $T_c$ above 300 K is itself due to the values of FPCs currently observed.

\subsection{Transitions between states of matter: solids, liquids and gases}
\label{firstor}

\begin{figure}
{\scalebox{0.4}{\includegraphics{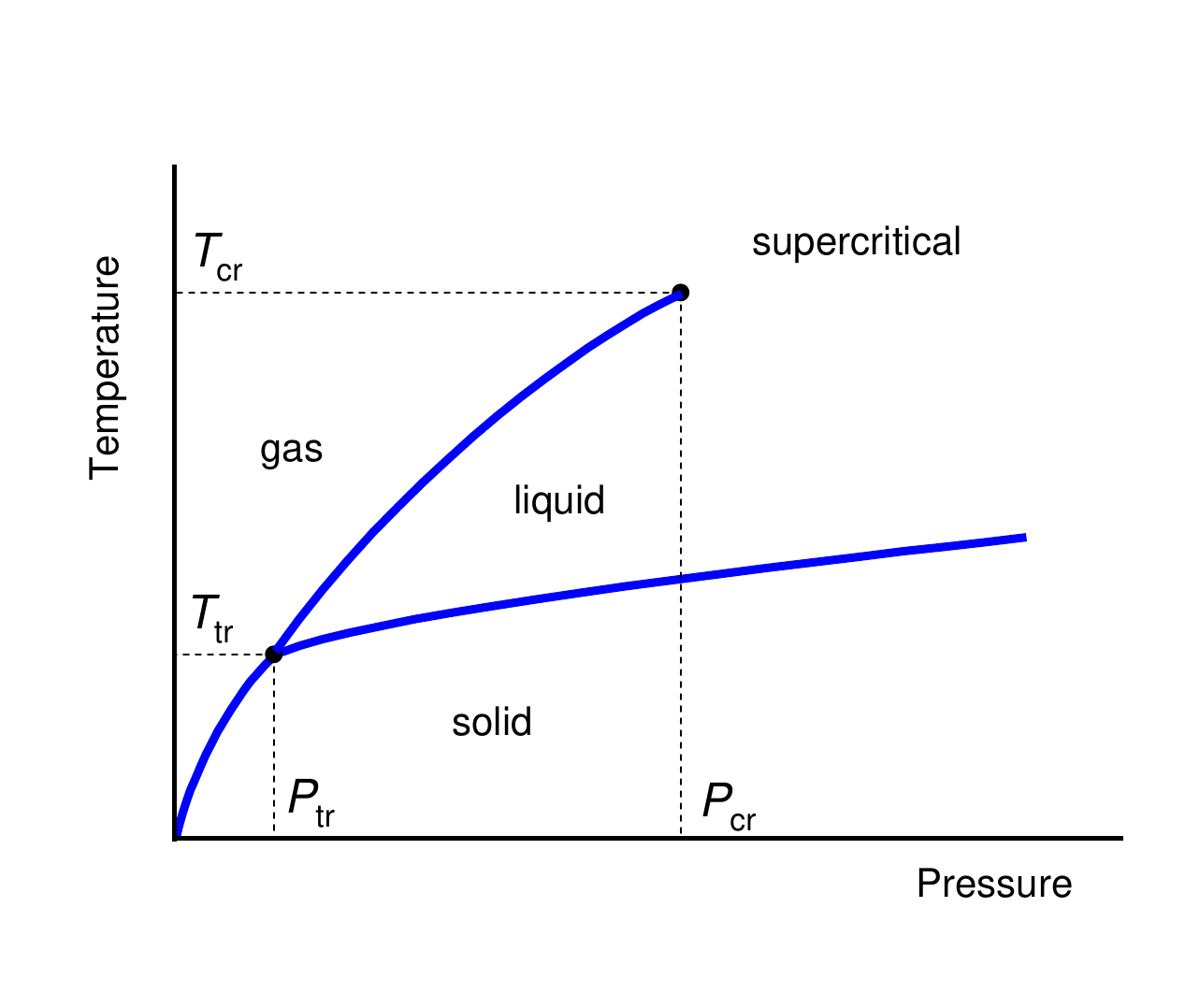}}}
\caption{Three states of matter (solids, liquids and gases) separated by three phase transition lines. Triple and critical points are shown.}
\label{fig3}
\end{figure}

Three basic states of matter, solids, liquids and gases, are represented by a phase diagram with three first-order phase transition lines as shown in Figure \ref{fig3}. In this section, we discuss how the observability of different states of matter and transitions between those states at our environmental conditions is related to FPCs.

The triple point pressure of many substances is below the atmospheric pressure, and so we often observe melting and boiling transitions on temperature increase. The phase co-existence lines in Figure \ref{fig3} are defined by the equality of thermodynamic potentials. At high pressure, melting temperature substantially depends on pressure (see Ref. \cite{premelting} for the functional form of pressure-temperature melting lines). At low pressure close to the triple point, melting can be discussed on the basis of cohesive energy \cite{ubbelohde} because microscopically, melting involves breaking chemical bonds. This cohesive energy is set by FPCs in Eq. \eqref{rydberg}. Melting temperatures correlate with cohesive energies and are on the order of $\omega_m$ or $\omega_{\rm D}$ in the range 10$^2$-10$^3$ K in many systems \cite{ubbelohde,grimvall-melting} (the difference between the cohesive energy and $\hbar\omega_m$ is due to the small factor $\left(\frac{m_e}{m}\right)^{\frac{1}{2}}$ in Eq. \eqref{ratio}). As discussed in Section \ref{secondor}, $\omega_m$ and $\omega_{\rm D}$ are set by FPCs. This gives the following role of FPCs in setting the states of matter.

FPCs resulting in large cohesive energies and melting temperatures imply that matter would exist in the solid state only at given environmental conditions. Melting and boiling phase transitions would be unobservable. Conversely, FPCs resulting in small cohesive energies and hence low melting and boiling temperatures imply that matter would exist in the gas state only.

We see that, similarly to structural phase transitions discussed in Section \ref{secondor}, the current values of FPCs set the observability of transitions between different states of matter in Figure \ref{fig3} and the existence of these states.

The apparent overlap of our environmental conditions with characteristic melting and boiling temperatures set by FPCs gives rise to two states important for life:

\begin{enumerate}
\item H$_2$O can exist as liquid water at our typical planetary conditions.
\item Carbon-based molecules and compounds can exist as cohesive (bonded) structures.
\end{enumerate}

Much larger cohesive energy due to different values of FPCs would result in H$_2$O existing as a solid only and hence no water-based life. Much smaller values of cohesive energies due to different FPCs would result in H$_2$O and carbon-based structures not existing as cohesive states but as gas states only.

A more detailed consideration shows an additional and more specific role of FPCs in life processes: FPCs have a biofriendly window constrained by biofriendly viscosity and diffusion setting the motion in essential processes in and across cells \cite{sciadv2023,ropp1}.


We note that classical entropies of solids and gases contain FPCs explicitly. Recalling the solid entropy $S_s\propto 3N\ln\frac{T}{\hbar\omega_{\rm D}}$ \cite{landaustat}, $\omega_{\rm D}\approx\omega_m$ and $\omega_m$ in Eq. \eqref{f1} with $m\propto m_p$ gives the dependence of $S_s$ on FPCs as

\begin{equation}
S_s\propto\ln\frac{\hbar^2m_p^{\frac{1}{2}}}{e^4m_e^{\frac{3}{2}}}
\label{ensf}
\end{equation}

The entropy of the classical gas is \cite{landaustat}

\begin{equation}
S_g=\frac{3}{2}N\left(\ln T+1\right)+N\left(\ln v+\ln\left(\frac{m}{2\pi\hbar^2}\right)^{\frac{3}{2}}+1\right)
\label{gasent}
\end{equation}

\noindent where $v$ is the volume per atom and $m$ is the atom mass.

Using $v\propto a_{\rm B}^3$, Eq. \eqref{bohr} and $m\propto m_p$ gives

\begin{equation}
S_g\propto\ln\frac{\hbar^3m_p^{\frac{3}{2}}}{e^6m_e^3}
\label{engf}
\end{equation}

The same FPCs interestingly feature in both numerators and denominators in Eqs. \eqref{ensf} and \eqref{engf}. $S_g$ in Eq. \eqref{engf} is more sensitive to FPCs as compared to $S_s$ in Eq.
\eqref{ensf}: larger $\hbar$ and smaller $m_e$ and $e$ increase $S_g$ faster. The same variation - larger $\hbar$ and smaller $m_e$ and $e$ - also decrease the cohesive energy in Eq. \eqref{rydberg},
favouring gases over solids. Hence gases may be viewed as more entropically favourable over solids from the FPCs perspective.

Similarly to what we saw in Section \ref{secondor} for structural phase transitions, the emerging high-temperature gas phase increases the entropy in comparison to what the entropy of the solid would
have been in the absence of phase transitions (for first-order transitions, the entropy increase has a contribution from the latent heat). Similarly to structural transitions, transitions between
different states of matter are constrained FPCs as discussed earlier in this section.


Differently from solid and gas entropies in Eqs. \eqref{soliden}, \eqref{entr3} and \eqref{gasent}, no equation for the liquid entropy exists due to the fundamental no-small parameter problem of the liquid theory (see Ref. \cite{premelting} for a quick review and Refs. \cite{ropp,mybook} for a more detailed discussion). Common models such as the hard-sphere and Van der Waals models and their extensions give the constant-volume specific heat $c_v=\frac{3}{2}$ and hence describe the ideal gas thermodynamically rather than real liquids where $c_v=3$ close to melting \cite{ropp,mybook}. However, the energy, the primary property in statistical physics \cite{landaustat}, {\it can} be calculated in liquids in the closed form using the main premise of statistical physics: the system's energy $E$ is set by the energies of its excitations \cite{landaustat1}. In liquids, these excitations are gapped phonons and particle jumps \cite{ropp,mybook,proctor1,proctor2,chen-review,withbook}. The entropy can then be found by integrating $\frac{c_v}{T}$, where $c_v=\frac{dE}{dT}$ (in almost entire range of liquid relaxation time, $E$ and $c_v$ are vibrational to a very good approximation \cite{duality}). Nevertheless, liquids are a cohesive state, and our discussion of how different states of matter emerge depending on cohesive energies set by FPCs applies to liquids too.


\subsection{Metastable states and chemical reactions}
\label{meta}

In the previous sections, we discussed phase transitions and the existence of different states as a result of these transitions. The usual statistical-mechanical treatment of these transitions assumes equilibrium phases corresponding to the global minimum of the free energy. However, many simple molecular compounds of light elements (e.g., most hydrocarbons, nitrogen oxides, hydrates, carbides, carbon oxide, alcohols, glycerin) are metastable and do not correspond to the lowest free energy minimum. Nevertheless, long-lived metastable phases in a certain range of pressure and temperature undergo reversible phase transitions which are consistent with equilibrium thermodynamics: the phase transition corresponds to the free energy crossover (see Ref. \cite{bra-meta} for review). When discussing transitions in these systems, the tacit assumption is the one we mentioned in Section \ref{secondor}: the observation time is assumed to be shorter than the relaxation time of the long-lived metastable state.

Going from elemental and simple substances to more complex compounds including organic compounds, we find that nearly all of them are metastable and do not correspond to the lowest free energy minimum for a given composition. The world of organic compounds relevant for life processes is nearly all metastable \cite{bra-meta}. Considering large inorganic structures suggests that the associated phase space can be astronomical if all possible conformations are considered (not accounting for the smaller number of actual viable existing structures) \cite{kauf-phase}.

Transitions between metastable states involve the same atomistic processes as transitions between equilibrium states. The kinetics of transitions
between metastable states is governed by the bond energies. These energies are governed by FPCs as discussed in Section \ref{observ}. Hence, similarly to equilibrium transitions, the observability and operation of transitions between different metastable states and the existence of numerous metastable phases is set by FPCs. The entropy of new high-temperature metastable states emerging in these transitions is expected to be higher because these states tend to be more symmetric with more open structure, as is the case for equilibrium transitons.

The same assertion of observability of transitions and different phases applies to chemical reactions and their products. Chemical reactions involve changes of chemical bonds with characteristic energies set by FPCs. Hence these reactions are governed by the cohesive energies set by FPCs and environmental parameters. For example, much larger cohesive energy due to different FPCs would imply that chemical reactions and their products (including those essential to life forms) would be inoperable at given environmental parameters.

\section{Fundamental constants and entropy increase}
\label{entropy}

We consider that understanding the current values of FPCs requires the discussion of their variation. Theories and observations of this variation were previously discussed \cite{smolin-bhs,barrow4alpha,seto-variation,spatialalpha,spatialalpha1,chiba,schellekens,martinsreview,barrow,adamsreview,uzan1,ropp1}. Anthropic arguments use this variation to rationalise the observed values of FPCs \cite{carrbook,finebook,adamsreview}. A different perspective should be noted: a final theory of most fundamental laws of nature, when found, would be rigid (``logically isolated'') \cite{weinberg-dreams} and guided by a unique design and an exquisite symmetry \cite{zeesymmetry}. FPCs would then be calculable from this theory.

In Section \ref{observ}, we discussed different cases where phase transitions and the emergence of different phases due to FPCs increase the entropy. We saw that different values of FPCs would
disable these transitions and different phases, resulting in smaller entropy. This suggests that the current values of FPCs give rise to {\it entropy increase}.

We discussed the entropy increase in condensed matter systems, however this increase applies to different length and energy scales in the Universe on general grounds. Recall Eq. \eqref{tcval} giving the temperature of phase transitions as a function of FPCs in condensed matter. However, the temperature of transitions and reactions depend on FPCs at other energy and length scales as well \cite{barrow,adamsreview}, enabling transitions and reactions at those scales. Similarly to condensed matter systems, we expect the new states emerging as a result of these transitions to have higher entropy. Hence we expect that FPCs promote entropy increase at different levels of structure in the Universe.

This entropy increase is specifically due to transitions and reactions and comes {\it in addition} to the conventional general mechanism of entropy increase (including in the second law of thermodynamics). In the conventional mechanism, $S$ of a closed system at a given energy $E_0$ is maximised at equilibrium because (a) the statistical probability $d\rho$ is

\begin{equation}
d\rho={\rm const}\cdot\delta(E-E_0)e^S\prod\limits_idE_i
\label{prob}
\end{equation}

\noindent where $E_i$ are energies of subsystems and (b) $d\rho$ is maximised when $E_i$ of all subsystems in their own equilibrium states are equal to their most probable equilibrium values
$\bar{E_i}$, corresponding to full equilibrium \cite{landaustat}. As discussed in Section \ref{secondor}, the subsystem where transitions and reactions take place (phase-changing subsystem, PCS) is
not closed and not in equilibrium because these transitions are enabled by the heat flow from another heat-supplying subsystem. However, subsystems can be identified which are (a) larger, include
both subsystems and {\it are} closed and (b) in equilibrium at long post-reaction times \cite{landaustat}. The conventional mechanism of increase of $S$ and $d\rho$ applies to these closed
subsystems, and our entropy increase in the PCS due to FPCs additionally increases the statistical probability in Eq. \eqref{prob}. As discussed below, identifying collections of subsystems which can
be considered as conditionally closed at different levels of the Universe can be of interest in future work.

We note that the environmental parameters can be variable in some processes. For example, nuclear synthesis reactions in a forming star depend on (environmental) pressure and temperature, which, in
turn, depend on the mass of collapsing interstellar gas. This variability can affect transitions and reactions and the properties of emergent structures. In some processes such as the above example
of nuclear synthesis in stars, the environmental parameters can also change as a result of reactions themselves. Nevertheless, the transitions and reactions are constrained by FPCs at varying
environmental parameters as they are at fixed parameters.

We recall the discussion of long-lived metastable states with higher entropy in Section \ref{observ}.
As recently reviewed \cite{bra-meta1}, metastable states are ubiquitous at different levels of matter organization in the Universe including nuclei and astronomical objects. The energy barriers
setting the lifetime of these states vary in a wide range 10$^{-7}$ to 10$^9$ eV per particle, with relaxation times often exceeding the age of the Universe \cite{bra-meta1}. These metastable states,
similarly to those discussed in Section \ref{observ}, emerge as a result of different transitions and reactions which depend on the values of FPCs \cite{barrow,adamsreview} and can co-exist at the
same environmental parameters.


The entropy increase due to FPCs can be discussed in terms of an evolution-like mechanism where this increase is analogous to selection in biological evolution due to environmental pressure. Based on
this entropy increase and the associated increase of statistical probability \eqref{prob}, we conjecture that total entropy increase is a {\it selection principle} for FPCs considered to vary in
earlier discussions (see, e.g., Refs. \cite{smolin-bhs,barrow4alpha,seto-variation,spatialalpha,spatialalpha1,chiba,schellekens,martinsreview,barrow,adamsreview,uzan1,ropp1} and references therein).
In other words, FPCs tend to give the most probable state. The mechanism of variability or fluctuations of FPCs, analogous to genetic mutation in biological evolution, is unknown, although several
proposals were discussed \cite{smolin-bhs,barrow4alpha,seto-variation,spatialalpha,spatialalpha1,chiba,schellekens,martinsreview,barrow,adamsreview,ropp1}. The above conjecture is nevertheless
interesting to consider because, if correct, it is analogous to introducing a Darwinian concept of a selection principle, the notion that brought new understanding even though the molecular DNA-based
mutation mechanism was not known at the time and came about a century later only.

Cyclic or oscillatory models of the Universe (see, e.g., Refs. \cite{cycle1,smolin-bhs,cycle11,cycle2,cycle3,cycle4,cycle5} and Ref. \cite{cycle-review} for a review including epistemic and philosophical aspects) can account for inheriting FPCs from one cycle to the next and maintain fluctuations around their values in the previous cycle. In these models, the change of FPCs is often related to the smallest size of the Universe and very high energy including the Planck energy scale where unknown physics operates. The size of the oscillating Universe at the inversion point in Ref. \cite{cycle5} is set by the system energy as a parameter in the model and does not need to be small.

We might ask whether the current FPCs give structures in the Universe that correspond to the maximised phase space and entropy or close to it. Adams concludes \cite{adamsreview} that the observable
Universe is not the most abundant and complex possible: somewhat different values of FPCs would give more complex and abundant structures (e.g. more baryons and stars, more efficient nuclear
synthesis and so on). Another interesting observation in Ref. \cite{adamsreview} is that our Universe does not lie at the centre of allowed range of values of FPCs. The evolutionary mechanism with
the entropy increase as a selection principle has several potential implications for these observations. We first note that this selection applies to FPCs and structures appearing due to these FPCs
existing at the time (as is the case in biological evolution) rather than to currently observed structures. Second, the observed structures depend on the magnitude of fluctuations of FPCs: large
fluctuations destroy the intricate network of constraints on the values of FPCs needed for different structures to exist \cite{adamsreview}, resulting in scarcity of structures and smaller phase
space. On the other hand, smaller fluctuations maintain the continuity of structures in different cycles. As a result, different structures can emerge along a sequence of cycles, with the entropy of
these structures fluctuating along the sequence. Third, it follows that maximising entropy as a function of FPCs is subject to many {\it constraints} on their values. Hence understanding the values
of FPCs (e.g. calculating these values or their ranges) invites an input from the optimisation theory. This programme would include taking an initial set of FPCs as variables, calculating the
entropies of important structures and formations in the Universe and finding FPCs or their ranges by maximising the total entropy while (a) accounting for constraints on FPCs from each structure
\cite{barrow,adamsreview} and (b) remembering that later-appearing structures may rely on the existence of earlier ones. This process prevents, for example, finding FPCs giving larger entropy of
earlier structures but smaller entropy of later structures (or no entropy of later structures if they are not formed due to unfavourable FPCs involved in the emergence of earlier structures) or FPCs
giving larger entropy of later structures but neglecting earlier structures needed for later structures to emerge.

An example related to this discussion and involving constraints is the black hole (BH) entropy. If most of the current entropy in the Universe is associated with the entropy of large BHs \cite{egan-bh}, one might consider small $\hbar$ and $c$ to maximise the BH entropy

\begin{equation}
S_{BH}=4\pi\frac{G}{\hbar c}M_{BH}^2
\end{equation}

\noindent where $M_{BH}$ is the BH mass. However, small $\hbar$ and $c$ give large $\alpha$ in Eq. \eqref{alpha}, breaking many constraints required for key processes of structure formation and
disrupting those processes \cite{barrow,adamsreview}. This includes processes preceding the formation of BHs such as stellar evolution which affect BHs at later times \cite{barrow,adamsreview}
(formation of large BHs is currently under discussion and includes the accretion involving smaller BHs).

The discussion of FPCs acting to increase the entropy and the conjecture of entropy increase as a selection principle for FPCs need further study. This includes detailing entropy changes in processes at different levels of structure organisation in the Universe \cite{barrow,adamsreview}, identifying sets of subsystems which can be considered closed or conditionally closed and where the entropy argument applies, clarifying the role of gravity \cite{landaustat} and so on. Given the scale of the problem we face in understanding the origin of fundamental constants \cite{barrow,weinberg,grandest}, making new observations and asking new questions is useful for progress.

\section{Discussion and summary}

Approaching the challenge to explain fundamental constants started with understanding the role these constants play. This role has historically been limited to high-energy nuclear physics and astrophysics physics and corresponding structures  \cite{barrow,barrow1,carr,carrbook,finebook,cahnreview,hoganreview,adamsreview,uzanreview}. Condensed matter physics was relatively unexplored in this regard, however more recently it was realised that FPCs set lower and upper bounds on key condensed matter properties. This has enriched condensed matter physics and deepened our understanding of its key phenomena.

Here, we discussed a much wider role played by FPCs in condensed matter physics: at given environmental parameters, FPCs define observability and operation of entire physical effects and phenomena. Our examples included the operation of different phase transitions and emerging new phases, existence of different states of matter, metastable phases, chemical reactions, their products and so on. These results were based on a theory of each physical effect and process rather than on a dimensional analysis as might have been construed. Ref. \cite{advphysreview} (pages 479-480) lists important differences between such a theory-based approach and a dimensional analysis.

We saw that were FPCs to take different values, transitions would be disabled at our environmental conditions and new different phases would not emerge. This suggests that the values of FPCs, by enabling transitions and new states emerging as a result of these transitions, act to increase the phase space and entropy.

The list of effects and phenomena in condensed matter physics discussed here can be probably extended significantly. The reason it is important to study how FPCs affect the operation and observability of different effects and phenomena in condensed matter is that this field is developed at the great level of detail. Physical models of condensed matter have been tested and cross-checked in numerous different experiments and simulations. In this sense, the insights into FPCs from condensed matter physics face less uncertainty and challenges as compared to models in high-energy physics and astrophysics (see Ref. \cite{adamsreview} for a review of these challenges).

If the insights into the role of FPCs in condensed matter physics are found to be general enough, these insights can be applied to structures at other levels in the Universe. An example is the insight that FPCs promote entropy increase, or the related conjecture of the selection principle for FPCs. These and similar insights can provide useful links between different areas of physics and open up new discussions.

Last but not least, deeper understanding the role of FPCs in condensed matter physics is important on its own. A realization that FPCs put lower or upper bounds on key condensed matter properties has
deepened our understanding of how these properties operate and benefited our physical models. These properties include viscosity \cite{sciadv}, diffusion and spin dynamics
\cite{spin,hbarm1,hbarm2,hbarm3,hbarm4}, electron and heat transport \cite{hartnoll,hartnoll1,behnia,nussinov1,grozdanov}, thermal conductivity in insulators \cite{momentumprb} and conductors
\cite{brareview}, the speed of sound \cite{sciadv2}, elastic moduli \cite{brareview,advphysreview}, flow in cells \cite{sciadv2023,ropp1}, vibration frequency in solids and superconducting
temperature \cite{ourtc}. Understanding how FPCs govern the observability and operation of entirely new effects and phenomena is another useful insight enhancing our understanding of condensed matter
physics.

I am grateful to  F. C. Adams, V. V. Brazhkin, M. T. Dove, M. Hutcheon, S. A. Kauffman, Z.-K. Liu, B. Monserrat, A. J. Phillips and C. J. Pickard for discussions and EPSRC for support.


\begin{thebibliography}{74}
\expandafter\ifx\csname natexlab\endcsname\relax\def\natexlab#1{#1}\fi \expandafter\ifx\csname bibnamefont\endcsname\relax
  \def\bibnamefont#1{#1}\fi
\expandafter\ifx\csname bibfnamefont\endcsname\relax
  \def\bibfnamefont#1{#1}\fi
\expandafter\ifx\csname citenamefont\endcsname\relax
  \def\citenamefont#1{#1}\fi
\expandafter\ifx\csname url\endcsname\relax
  \def\url#1{\texttt{#1}}\fi
\expandafter\ifx\csname urlprefix\endcsname\relax\def\urlprefix{URL }\fi \providecommand{\bibinfo}[2]{#2} \providecommand{\eprint}[2][]{\url{#2}}

\bibitem[{\citenamefont{Tiesinga et~al.}(2021)\citenamefont{Tiesinga, Mohr,
  Newell, and Taylor}}]{codata}
\bibinfo{author}{\bibfnamefont{E.}~\bibnamefont{Tiesinga}},
  \bibinfo{author}{\bibfnamefont{P.~J.} \bibnamefont{Mohr}},
  \bibinfo{author}{\bibfnamefont{D.~B.} \bibnamefont{Newell}},
  \bibnamefont{and} \bibinfo{author}{\bibfnamefont{B.~N.}
  \bibnamefont{Taylor}}, \bibinfo{journal}{Rev. Mod. Phys.}
  \textbf{\bibinfo{volume}{93}}, \bibinfo{pages}{025010}
  (\bibinfo{year}{2021}).

\bibitem[{\citenamefont{Ashcroft and Mermin}(1976)}]{ashcroft} \bibinfo{author}{\bibfnamefont{N.~W.} \bibnamefont{Ashcroft}} \bibnamefont{and}
  \bibinfo{author}{\bibfnamefont{N.~D.} \bibnamefont{Mermin}},
  \emph{\bibinfo{title}{Solid State Physics}} (\bibinfo{publisher}{Saunders
  College Publishing}, \bibinfo{year}{1976}).

\bibitem[{\citenamefont{Uzan}(2003)}]{uzanreview} \bibinfo{author}{\bibfnamefont{J.~P.} \bibnamefont{Uzan}},
  \bibinfo{journal}{Rev. Mod. Phys.} \textbf{\bibinfo{volume}{75}},
  \bibinfo{pages}{403} (\bibinfo{year}{2003}).

\bibitem[{\citenamefont{Uzan}(2011)}]{uzan1} \bibinfo{author}{\bibfnamefont{J.~P.} \bibnamefont{Uzan}},
  \bibinfo{journal}{Living Rev. Relativity} \textbf{\bibinfo{volume}{14}},
  \bibinfo{pages}{2} (\bibinfo{year}{2011}).

\bibitem[{\citenamefont{NIST}()}]{nist-fund} \bibinfo{author}{\bibnamefont{NIST}}, \emph{\bibinfo{title}{Fundamental
  Physical Constants, https://physics.nist.gov/cuu/Constants}} (????).

\bibitem[{\citenamefont{Barrow}(2003)}]{barrow} \bibinfo{author}{\bibfnamefont{J.~D.} \bibnamefont{Barrow}},
  \emph{\bibinfo{title}{The Constants of Nature}} (\bibinfo{publisher}{Pantheon
  Books}, \bibinfo{year}{2003}).

\bibitem[{\citenamefont{Barrow and Tipler}(2009)}]{barrow1} \bibinfo{author}{\bibfnamefont{J.~D.} \bibnamefont{Barrow}} \bibnamefont{and}
  \bibinfo{author}{\bibfnamefont{F.~J.} \bibnamefont{Tipler}},
  \emph{\bibinfo{title}{The Anthropic Cosmological Principle}}
  (\bibinfo{publisher}{Oxford University Press}, \bibinfo{year}{2009}).

\bibitem[{\citenamefont{Carr and Rees}(1979)}]{carr} \bibinfo{author}{\bibfnamefont{B.~J.} \bibnamefont{Carr}} \bibnamefont{and}
  \bibinfo{author}{\bibfnamefont{M.~J.} \bibnamefont{Rees}},
  \bibinfo{journal}{Nature} \textbf{\bibinfo{volume}{278}},
  \bibinfo{pages}{605} (\bibinfo{year}{1979}).

\bibitem[{\citenamefont{Carr}(2009)}]{carrbook} \bibinfo{author}{\bibfnamefont{B.}~\bibnamefont{Carr}},
  \emph{\bibinfo{title}{Universe or Multiverse?}}
  (\bibinfo{publisher}{Cambridge University Press}, \bibinfo{year}{2009}).

\bibitem[{\citenamefont{Sloan et~al.}(2022)\citenamefont{Sloan, Batista, Hicks,
  and Davies}}]{finebook}
\bibinfo{editor}{\bibfnamefont{D.}~\bibnamefont{Sloan}},
  \bibinfo{editor}{\bibfnamefont{R.~A.} \bibnamefont{Batista}},
  \bibinfo{editor}{\bibfnamefont{M.~T.} \bibnamefont{Hicks}}, \bibnamefont{and}
  \bibinfo{editor}{\bibfnamefont{R.}~\bibnamefont{Davies}}, eds.,
  \emph{\bibinfo{title}{Fine Tuning in the Physical Universe}}
  (\bibinfo{publisher}{Cambridge University Press}, \bibinfo{year}{2022}).

\bibitem[{\citenamefont{Cahn}(1996)}]{cahnreview} \bibinfo{author}{\bibfnamefont{R.~N.} \bibnamefont{Cahn}},
  \bibinfo{journal}{Rev. Mod. Phys.} \textbf{\bibinfo{volume}{68}},
  \bibinfo{pages}{951} (\bibinfo{year}{1996}).

\bibitem[{\citenamefont{Hogan}(2000)}]{hoganreview} \bibinfo{author}{\bibfnamefont{C.~J.} \bibnamefont{Hogan}},
  \bibinfo{journal}{Rev. Mod. Phys.} \textbf{\bibinfo{volume}{72}},
  \bibinfo{pages}{1149} (\bibinfo{year}{2000}).

\bibitem[{\citenamefont{Adams}(2019)}]{adamsreview} \bibinfo{author}{\bibfnamefont{F.~C.} \bibnamefont{Adams}},
  \bibinfo{journal}{Physics Reports} \textbf{\bibinfo{volume}{807}},
  \bibinfo{pages}{1} (\bibinfo{year}{2019}).

\bibitem[{\citenamefont{Barrow and Webb}(2006)}]{grandest} \bibinfo{author}{\bibfnamefont{J.~D.} \bibnamefont{Barrow}} \bibnamefont{and}
  \bibinfo{author}{\bibfnamefont{J.~K.} \bibnamefont{Webb}},
  \bibinfo{journal}{Sci. Amer.} \textbf{\bibinfo{volume}{16}},
  \bibinfo{pages}{64} (\bibinfo{year}{2006}).

\bibitem[{\citenamefont{Weinberg}(1983)}]{weinberg} \bibinfo{author}{\bibfnamefont{S.}~\bibnamefont{Weinberg}},
  \bibinfo{journal}{Phil. Trans. R. Soc. Lond. A}
  \textbf{\bibinfo{volume}{310}}, \bibinfo{pages}{249} (\bibinfo{year}{1983}).

\bibitem[{\citenamefont{Hogan}(2009)}]{hoganbook} \bibinfo{author}{\bibfnamefont{C.~J.} \bibnamefont{Hogan}}, in
  \emph{\bibinfo{booktitle}{Universe or Multiverse?}}, edited by
  \bibinfo{editor}{\bibfnamefont{B.}~\bibnamefont{Carr}}
  (\bibinfo{publisher}{Cambridge University Press}, \bibinfo{year}{2009}), p.
  \bibinfo{pages}{221}.

\bibitem[{\citenamefont{Trachenko and Brazhkin}(2020)}]{sciadv} \bibinfo{author}{\bibfnamefont{K.}~\bibnamefont{Trachenko}} \bibnamefont{and}
  \bibinfo{author}{\bibfnamefont{V.~V.} \bibnamefont{Brazhkin}},
  \bibinfo{journal}{Sci. Adv.} \textbf{\bibinfo{volume}{6}},
  \bibinfo{pages}{eaba3747} (\bibinfo{year}{2020}).

\bibitem[{\citenamefont{Luciuk et~al.}(2017)}]{spin} \bibinfo{author}{\bibfnamefont{C.}~\bibnamefont{Luciuk}} \bibnamefont{et~al.},
  \bibinfo{journal}{Phys. Rev. Lett.} \textbf{\bibinfo{volume}{118}},
  \bibinfo{pages}{130405} (\bibinfo{year}{2017}).

\bibitem[{\citenamefont{Sommer et~al.}(2011)\citenamefont{Sommer, Ku, Roati,
  and Zwierlein}}]{hbarm1}
\bibinfo{author}{\bibfnamefont{A.}~\bibnamefont{Sommer}},
  \bibinfo{author}{\bibfnamefont{M.}~\bibnamefont{Ku}},
  \bibinfo{author}{\bibfnamefont{G.}~\bibnamefont{Roati}}, \bibnamefont{and}
  \bibinfo{author}{\bibnamefont{Zwierlein}}, \bibinfo{journal}{Nature}
  \textbf{\bibinfo{volume}{472}}, \bibinfo{pages}{201} (\bibinfo{year}{2011}).

\bibitem[{\citenamefont{Bardon et~al.}(2014)}]{hbarm2} \bibinfo{author}{\bibfnamefont{A.~B.} \bibnamefont{Bardon}}
  \bibnamefont{et~al.}, \bibinfo{journal}{Science}
  \textbf{\bibinfo{volume}{344}}, \bibinfo{pages}{722} (\bibinfo{year}{2014}).

\bibitem[{\citenamefont{Trotzky et~al.}(2015)}]{hbarm3} \bibinfo{author}{\bibfnamefont{S.}~\bibnamefont{Trotzky}} \bibnamefont{et~al.},
  \bibinfo{journal}{Phys. Rev. Lett.} \textbf{\bibinfo{volume}{114}},
  \bibinfo{pages}{015301} (\bibinfo{year}{2015}).

\bibitem[{\citenamefont{Enss and Thywissen}(2018)}]{hbarm4} \bibinfo{author}{\bibfnamefont{T.}~\bibnamefont{Enss}} \bibnamefont{and}
  \bibinfo{author}{\bibfnamefont{J.~H.} \bibnamefont{Thywissen}},
  \bibinfo{journal}{Ann. Rev. Condens. Matter Phys.}
  \textbf{\bibinfo{volume}{10}}, \bibinfo{pages}{85} (\bibinfo{year}{2018}).

\bibitem[{\citenamefont{Hartnoll}(2015)}]{hartnoll} \bibinfo{author}{\bibfnamefont{S.~A.} \bibnamefont{Hartnoll}},
  \bibinfo{journal}{Nat. Phys.} \textbf{\bibinfo{volume}{11}},
  \bibinfo{pages}{54} (\bibinfo{year}{2015}).

\bibitem[{\citenamefont{Mousatov and Hartnoll}(2020)}]{hartnoll1} \bibinfo{author}{\bibfnamefont{H.}~\bibnamefont{Mousatov}} \bibnamefont{and}
  \bibinfo{author}{\bibfnamefont{S.~A.} \bibnamefont{Hartnoll}},
  \bibinfo{journal}{Nat. Phys.} \textbf{\bibinfo{volume}{16}},
  \bibinfo{pages}{579} (\bibinfo{year}{2020}).

\bibitem[{\citenamefont{Behnia and Kapitulnik}(2019)}]{behnia} \bibinfo{author}{\bibfnamefont{K.}~\bibnamefont{Behnia}} \bibnamefont{and}
  \bibinfo{author}{\bibfnamefont{A.}~\bibnamefont{Kapitulnik}},
  \bibinfo{journal}{J. Phys.: Condens. Matt.} \textbf{\bibinfo{volume}{31}},
  \bibinfo{pages}{405702} (\bibinfo{year}{2019}).

\bibitem[{\citenamefont{Nussinov and Chakrabarty}(2022)}]{nussinov1} \bibinfo{author}{\bibfnamefont{Z.}~\bibnamefont{Nussinov}} \bibnamefont{and}
  \bibinfo{author}{\bibfnamefont{S.}~\bibnamefont{Chakrabarty}},
  \bibinfo{journal}{Annals of Physics} \textbf{\bibinfo{volume}{443}},
  \bibinfo{pages}{168970} (\bibinfo{year}{2022}).

\bibitem[{\citenamefont{Grozdanov}(2021)}]{grozdanov} \bibinfo{author}{\bibfnamefont{S.}~\bibnamefont{Grozdanov}},
  \bibinfo{journal}{Phys. Rev. Lett.} \textbf{\bibinfo{volume}{126}},
  \bibinfo{pages}{051601} (\bibinfo{year}{2021}).

\bibitem[{\citenamefont{Trachenko et~al.}(2021)\citenamefont{Trachenko,
  Baggioli, Behnia, and Brazhkin}}]{momentumprb}
\bibinfo{author}{\bibfnamefont{K.}~\bibnamefont{Trachenko}},
  \bibinfo{author}{\bibfnamefont{M.}~\bibnamefont{Baggioli}},
  \bibinfo{author}{\bibfnamefont{K.}~\bibnamefont{Behnia}}, \bibnamefont{and}
  \bibinfo{author}{\bibfnamefont{V.~V.} \bibnamefont{Brazhkin}},
  \bibinfo{journal}{Phys. Rev. B} \textbf{\bibinfo{volume}{103}},
  \bibinfo{pages}{014311} (\bibinfo{year}{2021}).

\bibitem[{\citenamefont{Brazhkin}(2023)}]{brareview} \bibinfo{author}{\bibfnamefont{V.~V.} \bibnamefont{Brazhkin}},
  \bibinfo{journal}{Phys.-Usp.} \textbf{\bibinfo{volume}{66}},
  \bibinfo{pages}{1154} (\bibinfo{year}{2023}).

\bibitem[{\citenamefont{Trachenko et~al.}(2020)\citenamefont{Trachenko,
  Monserrat, Pickard, and Brazhkin}}]{sciadv2}
\bibinfo{author}{\bibfnamefont{K.}~\bibnamefont{Trachenko}},
  \bibinfo{author}{\bibfnamefont{B.}~\bibnamefont{Monserrat}},
  \bibinfo{author}{\bibfnamefont{C.~J.} \bibnamefont{Pickard}},
  \bibnamefont{and} \bibinfo{author}{\bibfnamefont{V.~V.}
  \bibnamefont{Brazhkin}}, \bibinfo{journal}{Sci. Adv.}
  \textbf{\bibinfo{volume}{6}}, \bibinfo{pages}{eabc8662}
  (\bibinfo{year}{2020}).

\bibitem[{\citenamefont{Trachenko}(2023{\natexlab{a}})}]{advphysreview} \bibinfo{author}{\bibfnamefont{K.}~\bibnamefont{Trachenko}},
  \bibinfo{journal}{Advances in Physics} \textbf{\bibinfo{volume}{70}},
  \bibinfo{pages}{469} (\bibinfo{year}{2023}{\natexlab{a}}).

\bibitem[{\citenamefont{Trachenko}(2023{\natexlab{b}})}]{sciadv2023} \bibinfo{author}{\bibfnamefont{K.}~\bibnamefont{Trachenko}},
  \bibinfo{journal}{Science Adv.} \textbf{\bibinfo{volume}{9}},
  \bibinfo{pages}{eadh9024} (\bibinfo{year}{2023}{\natexlab{b}}).

\bibitem[{\citenamefont{Trachenko}(2023{\natexlab{c}})}]{ropp1} \bibinfo{author}{\bibfnamefont{K.}~\bibnamefont{Trachenko}},
  \bibinfo{journal}{Rep. Prog. Phys.} \textbf{\bibinfo{volume}{86}},
  \bibinfo{pages}{112601} (\bibinfo{year}{2023}{\natexlab{c}}).

\bibitem[{\citenamefont{Trachenko et~al.}()\citenamefont{Trachenko, Monserrat,
  Hutcheon, and Pickard}}]{ourtc}
\bibinfo{author}{\bibfnamefont{K.}~\bibnamefont{Trachenko}},
  \bibinfo{author}{\bibfnamefont{B.}~\bibnamefont{Monserrat}},
  \bibinfo{author}{\bibfnamefont{M.}~\bibnamefont{Hutcheon}}, \bibnamefont{and}
  \bibinfo{author}{\bibfnamefont{C.~J.} \bibnamefont{Pickard}},
  \bibinfo{journal}{arxiv:2406.08129}  (????).

\bibitem[{\citenamefont{Weinberg}(1993)}]{weinberg-dreams} \bibinfo{author}{\bibfnamefont{S.}~\bibnamefont{Weinberg}},
  \emph{\bibinfo{title}{Dreams of a final theory}} (\bibinfo{publisher}{Vintage
  books}, \bibinfo{year}{1993}).

\bibitem[{\citenamefont{Hammonds et~al.}(1996)\citenamefont{Hammonds, Dove,
  Giddy, Heine, and Winkler}}]{dove-rums1}
\bibinfo{author}{\bibfnamefont{K.~D.} \bibnamefont{Hammonds}},
  \bibinfo{author}{\bibfnamefont{M.~T.} \bibnamefont{Dove}},
  \bibinfo{author}{\bibfnamefont{A.~P.} \bibnamefont{Giddy}},
  \bibinfo{author}{\bibfnamefont{V.}~\bibnamefont{Heine}}, \bibnamefont{and}
  \bibinfo{author}{\bibfnamefont{B.}~\bibnamefont{Winkler}},
  \bibinfo{journal}{Amer. Mineral.} \textbf{\bibinfo{volume}{81}},
  \bibinfo{pages}{1057} (\bibinfo{year}{1996}).

\bibitem[{\citenamefont{Dove}(1997)}]{dove-rums2} \bibinfo{author}{\bibfnamefont{M.~T.} \bibnamefont{Dove}},
  \bibinfo{journal}{Amer. Mineral.} \textbf{\bibinfo{volume}{82}},
  \bibinfo{pages}{213} (\bibinfo{year}{1997}).

\bibitem[{\citenamefont{Landau and Lifshitz}(1970)}]{landaustat} \bibinfo{author}{\bibfnamefont{L.~D.} \bibnamefont{Landau}} \bibnamefont{and}
  \bibinfo{author}{\bibfnamefont{E.~M.} \bibnamefont{Lifshitz}},
  \emph{\bibinfo{title}{Course of Theoretical Physics, vol. 5. Statistical
  Physics, part 1.}} (\bibinfo{publisher}{Pergamon Press},
  \bibinfo{year}{1970}).

\bibitem[{\citenamefont{Dove}(2003)}]{dovesecbook} \bibinfo{author}{\bibfnamefont{M.~T.} \bibnamefont{Dove}},
  \emph{\bibinfo{title}{Structure and Dynamics– An Atomic View of Materials}}
  (\bibinfo{publisher}{Oxford University Press}, \bibinfo{year}{2003}).

\bibitem[{\citenamefont{Dove}(1993)}]{doveblue} \bibinfo{author}{\bibfnamefont{M.~T.} \bibnamefont{Dove}},
  \emph{\bibinfo{title}{Introduction to Lattice Dynamics}}
  (\bibinfo{publisher}{Cambridge University Press}, \bibinfo{year}{1993}).

\bibitem[{\citenamefont{Flores-Livas et~al.}(2020)\citenamefont{Flores-Livas,
  Boeri, Sanna, Profeta, Arita, and Eremets}}]{hreview1}
\bibinfo{author}{\bibfnamefont{J.~A.} \bibnamefont{Flores-Livas}},
  \bibinfo{author}{\bibfnamefont{L.}~\bibnamefont{Boeri}},
  \bibinfo{author}{\bibfnamefont{A.}~\bibnamefont{Sanna}},
  \bibinfo{author}{\bibfnamefont{G.}~\bibnamefont{Profeta}},
  \bibinfo{author}{\bibfnamefont{R.}~\bibnamefont{Arita}}, \bibnamefont{and}
  \bibinfo{author}{\bibfnamefont{M.}~\bibnamefont{Eremets}},
  \bibinfo{journal}{Phys. Rep.} \textbf{\bibinfo{volume}{856}},
  \bibinfo{pages}{1} (\bibinfo{year}{2020}).

\bibitem[{\citenamefont{Duan et~al.}(2017)\citenamefont{Duan, Liu, Ma, Shao,
  Liu, and Cui}}]{hreview2}
\bibinfo{author}{\bibfnamefont{D.}~\bibnamefont{Duan}},
  \bibinfo{author}{\bibfnamefont{Y.}~\bibnamefont{Liu}},
  \bibinfo{author}{\bibfnamefont{Y.}~\bibnamefont{Ma}},
  \bibinfo{author}{\bibfnamefont{Z.}~\bibnamefont{Shao}},
  \bibinfo{author}{\bibfnamefont{B.}~\bibnamefont{Liu}}, \bibnamefont{and}
  \bibinfo{author}{\bibfnamefont{T.}~\bibnamefont{Cui}},
  \bibinfo{journal}{Natl. Sci. Rev.} \textbf{\bibinfo{volume}{4}},
  \bibinfo{pages}{121} (\bibinfo{year}{2017}).

\bibitem[{\citenamefont{Pickard et~al.}(2019)\citenamefont{Pickard, Errea, and
  Eremets}}]{hreview3}
\bibinfo{author}{\bibfnamefont{C.~J.} \bibnamefont{Pickard}},
  \bibinfo{author}{\bibfnamefont{I.}~\bibnamefont{Errea}}, \bibnamefont{and}
  \bibinfo{author}{\bibfnamefont{M.~I.} \bibnamefont{Eremets}},
  \bibinfo{journal}{Ann. Rev. Condens. Matt. Phys.}
  \textbf{\bibinfo{volume}{11}}, \bibinfo{pages}{57} (\bibinfo{year}{2019}).

\bibitem[{\citenamefont{Trachenko}(2024)}]{premelting} \bibinfo{author}{\bibfnamefont{K.}~\bibnamefont{Trachenko}},
  \bibinfo{journal}{Phys. Rev. E} \textbf{\bibinfo{volume}{109}},
  \bibinfo{pages}{034122} (\bibinfo{year}{2024}).

\bibitem[{\citenamefont{Ubbelohde}(1978)}]{ubbelohde} \bibinfo{author}{\bibfnamefont{A.~R.} \bibnamefont{Ubbelohde}},
  \emph{\bibinfo{title}{The molten state of matter}} (\bibinfo{publisher}{J.
  Wiley and Sons}, \bibinfo{year}{1978}).

\bibitem[{\citenamefont{Grimvall and Sj\"{o}din}(1974)}]{grimvall-melting} \bibinfo{author}{\bibfnamefont{G.}~\bibnamefont{Grimvall}} \bibnamefont{and}
  \bibinfo{author}{\bibnamefont{Sj\"{o}din}}, \bibinfo{journal}{Phys. Scripta}
  \textbf{\bibinfo{volume}{10}}, \bibinfo{pages}{340} (\bibinfo{year}{1974}).

\bibitem[{\citenamefont{Trachenko and Brazhkin}(2016)}]{ropp} \bibinfo{author}{\bibfnamefont{K.}~\bibnamefont{Trachenko}} \bibnamefont{and}
  \bibinfo{author}{\bibfnamefont{V.~V.} \bibnamefont{Brazhkin}},
  \bibinfo{journal}{Rep. Prog. Phys.} \textbf{\bibinfo{volume}{79}},
  \bibinfo{pages}{016502} (\bibinfo{year}{2016}).

\bibitem[{\citenamefont{Trachenko}(2023{\natexlab{d}})}]{mybook} \bibinfo{author}{\bibfnamefont{K.}~\bibnamefont{Trachenko}},
  \emph{\bibinfo{title}{Theory of Liquids: from Excitations to Thermodynamics}}
  (\bibinfo{publisher}{Cambridge University Press},
  \bibinfo{year}{2023}{\natexlab{d}}).

\bibitem[{\citenamefont{Lifshitz and Pitaevskii}(2006)}]{landaustat1} \bibinfo{author}{\bibfnamefont{E.~M.} \bibnamefont{Lifshitz}} \bibnamefont{and}
  \bibinfo{author}{\bibfnamefont{L.~P.} \bibnamefont{Pitaevskii}},
  \emph{\bibinfo{title}{Course of Theoretical Physics, vol.9. Statistical
  Physics, Part 2}} (\bibinfo{publisher}{Pergamon Press},
  \bibinfo{year}{2006}).

\bibitem[{\citenamefont{Proctor}(2020)}]{proctor1} \bibinfo{author}{\bibfnamefont{J.}~\bibnamefont{Proctor}},
  \bibinfo{journal}{Phys. Fluids} \textbf{\bibinfo{volume}{32}},
  \bibinfo{pages}{107105} (\bibinfo{year}{2020}).

\bibitem[{\citenamefont{Proctor}(2021)}]{proctor2} \bibinfo{author}{\bibfnamefont{J.}~\bibnamefont{Proctor}},
  \emph{\bibinfo{title}{The liquid and supercritical states of matter}}
  (\bibinfo{publisher}{CRC Press}, \bibinfo{year}{2021}).

\bibitem[{\citenamefont{Chen}(2022)}]{chen-review} \bibinfo{author}{\bibfnamefont{G.}~\bibnamefont{Chen}},
  \bibinfo{journal}{Journal of Heat Transfer} \textbf{\bibinfo{volume}{144}},
  \bibinfo{pages}{010801} (\bibinfo{year}{2022}).

\bibitem[{\citenamefont{de~With}(2024)}]{withbook} \bibinfo{author}{\bibfnamefont{G.}~\bibnamefont{de~With}},
  \emph{\bibinfo{title}{Phases of Matter and their Transitions: Concepts and
  Principles for Chemists, Physicists, Engineers, and Materials Scientists}}
  (\bibinfo{publisher}{Wiley}, \bibinfo{year}{2024}).

\bibitem[{\citenamefont{Trachenko and Brazhkin}(2013)}]{duality} \bibinfo{author}{\bibfnamefont{K.}~\bibnamefont{Trachenko}} \bibnamefont{and}
  \bibinfo{author}{\bibfnamefont{V.~V.} \bibnamefont{Brazhkin}},
  \bibinfo{journal}{Sci. Rep.} \textbf{\bibinfo{volume}{3}},
  \bibinfo{pages}{2188} (\bibinfo{year}{2013}).

\bibitem[{\citenamefont{Brazhkin}(2006)}]{bra-meta} \bibinfo{author}{\bibfnamefont{V.~V.} \bibnamefont{Brazhkin}},
  \bibinfo{journal}{J. Phys.: Condens. Matt} \textbf{\bibinfo{volume}{18}},
  \bibinfo{pages}{9643} (\bibinfo{year}{2006}).

\bibitem[{\citenamefont{Cortes et~al.}()\citenamefont{Cortes, Kauffman, Liddle,
  and Smolin}}]{kauf-phase}
\bibinfo{author}{\bibfnamefont{M.}~\bibnamefont{Cortes}},
  \bibinfo{author}{\bibfnamefont{S.~A.} \bibnamefont{Kauffman}},
  \bibinfo{author}{\bibfnamefont{A.~R.} \bibnamefont{Liddle}},
  \bibnamefont{and} \bibinfo{author}{\bibfnamefont{L.}~\bibnamefont{Smolin}},
  \bibinfo{journal}{arxiv:2204.09378}  (????).

\bibitem[{\citenamefont{Smolin}(1992)}]{smolin-bhs} \bibinfo{author}{\bibfnamefont{L.}~\bibnamefont{Smolin}},
  \bibinfo{journal}{Class. Quantum Grav.} \textbf{\bibinfo{volume}{9}},
  \bibinfo{pages}{173} (\bibinfo{year}{1992}).

\bibitem[{\citenamefont{Wilczynska et~al.}(2020)}]{barrow4alpha} \bibinfo{author}{\bibfnamefont{M.~R.} \bibnamefont{Wilczynska}}
  \bibnamefont{et~al.}, \bibinfo{journal}{Sci. Adv.}
  \textbf{\bibinfo{volume}{6}}, \bibinfo{pages}{eaay9672}
  (\bibinfo{year}{2020}).

\bibitem[{\citenamefont{Seto et~al.}(2023)\citenamefont{Seto, Takahashi, and
  Y.}}]{seto-variation}
\bibinfo{author}{\bibfnamefont{O.}~\bibnamefont{Seto}},
  \bibinfo{author}{\bibfnamefont{T.}~\bibnamefont{Takahashi}},
  \bibnamefont{and} \bibinfo{author}{\bibfnamefont{T.}~\bibnamefont{Y.}},
  \bibinfo{journal}{Phys. Rev. D} \textbf{\bibinfo{volume}{108}},
  \bibinfo{pages}{023525} (\bibinfo{year}{2023}).

\bibitem[{\citenamefont{Webb et~al.}(2011)\citenamefont{Webb, King, Murphy,
  Flambaum, Carswell, and Bainbridge}}]{spatialalpha}
\bibinfo{author}{\bibfnamefont{J.~K.} \bibnamefont{Webb}},
  \bibinfo{author}{\bibfnamefont{J.~A.} \bibnamefont{King}},
  \bibinfo{author}{\bibfnamefont{M.~T.} \bibnamefont{Murphy}},
  \bibinfo{author}{\bibfnamefont{V.~V.} \bibnamefont{Flambaum}},
  \bibinfo{author}{\bibfnamefont{R.~F.} \bibnamefont{Carswell}},
  \bibnamefont{and} \bibinfo{author}{\bibfnamefont{M.~B.}
  \bibnamefont{Bainbridge}}, \bibinfo{journal}{Phys. Rev. Lett.}
  \textbf{\bibinfo{volume}{107}}, \bibinfo{pages}{191101}
  (\bibinfo{year}{2011}).

\bibitem[{\citenamefont{King et~al.}(2012)}]{spatialalpha1} \bibinfo{author}{\bibfnamefont{J.~A.} \bibnamefont{King}} \bibnamefont{et~al.},
  \bibinfo{journal}{Mon. Not. R. Astron. Soc.} \textbf{\bibinfo{volume}{422}},
  \bibinfo{pages}{3370} (\bibinfo{year}{2012}).

\bibitem[{\citenamefont{Chiba}(2011)}]{chiba} \bibinfo{author}{\bibfnamefont{T.}~\bibnamefont{Chiba}},
  \bibinfo{journal}{Prog. Theor. Phys.} \textbf{\bibinfo{volume}{126}},
  \bibinfo{pages}{993} (\bibinfo{year}{2011}).

\bibitem[{\citenamefont{Schellekens}(2013)}]{schellekens} \bibinfo{author}{\bibfnamefont{A.~N.} \bibnamefont{Schellekens}},
  \bibinfo{journal}{Rev. Mod. Phys.} \textbf{\bibinfo{volume}{85}},
  \bibinfo{pages}{1491} (\bibinfo{year}{2013}).

\bibitem[{\citenamefont{Martins}(2017)}]{martinsreview} \bibinfo{author}{\bibfnamefont{C.~J. A.~P.} \bibnamefont{Martins}},
  \bibinfo{journal}{Rep. Prog. Phys.} \textbf{\bibinfo{volume}{80}},
  \bibinfo{pages}{126902} (\bibinfo{year}{2017}).

\bibitem[{\citenamefont{Zee}(2007)}]{zeesymmetry} \bibinfo{author}{\bibfnamefont{A.}~\bibnamefont{Zee}},
  \emph{\bibinfo{title}{Fearful Symmetry}} (\bibinfo{publisher}{Princeton
  University Press}, \bibinfo{year}{2007}).

\bibitem[{\citenamefont{Brazhkin}(2024)}]{bra-meta1} \bibinfo{author}{\bibfnamefont{V.~V.} \bibnamefont{Brazhkin}},
  \bibinfo{journal}{JETP. Lett.} \textbf{\bibinfo{volume}{119}},
  \bibinfo{pages}{972} (\bibinfo{year}{2024}).

\bibitem[{\citenamefont{Adams and Laughlin}(1997)}]{cycle1} \bibinfo{author}{\bibfnamefont{F.~C.} \bibnamefont{Adams}} \bibnamefont{and}
  \bibinfo{author}{\bibfnamefont{G.}~\bibnamefont{Laughlin}},
  \bibinfo{journal}{Rev. Mod. Phys.} \textbf{\bibinfo{volume}{69}},
  \bibinfo{pages}{337} (\bibinfo{year}{1997}).

\bibitem[{\citenamefont{Steinhardt and Turok}(2002)}]{cycle11} \bibinfo{author}{\bibfnamefont{P.~S.} \bibnamefont{Steinhardt}}
  \bibnamefont{and} \bibinfo{author}{\bibfnamefont{N.}~\bibnamefont{Turok}},
  \bibinfo{journal}{Phys. Rev. D} \textbf{\bibinfo{volume}{65}},
  \bibinfo{pages}{126003} (\bibinfo{year}{2002}).

\bibitem[{\citenamefont{Baum and Frampton}(2007)}]{cycle2} \bibinfo{author}{\bibfnamefont{L.}~\bibnamefont{Baum}} \bibnamefont{and}
  \bibinfo{author}{\bibfnamefont{P.~H.} \bibnamefont{Frampton}},
  \bibinfo{journal}{Phys. Rev. Lett.} \textbf{\bibinfo{volume}{98}},
  \bibinfo{pages}{071301} (\bibinfo{year}{2007}).

\bibitem[{\citenamefont{Ashtekar and Singh}(2011)}]{cycle3} \bibinfo{author}{\bibfnamefont{A.}~\bibnamefont{Ashtekar}} \bibnamefont{and}
  \bibinfo{author}{\bibfnamefont{P.}~\bibnamefont{Singh}},
  \bibinfo{journal}{Class. Quantum Grav.} \textbf{\bibinfo{volume}{28}},
  \bibinfo{pages}{213001} (\bibinfo{year}{2011}).

\bibitem[{\citenamefont{Gurzadyan and Penrose}(2013)}]{cycle4} \bibinfo{author}{\bibfnamefont{V.~G.} \bibnamefont{Gurzadyan}}
  \bibnamefont{and} \bibinfo{author}{\bibfnamefont{R.}~\bibnamefont{Penrose}},
  \bibinfo{journal}{Eur. Phys. J. Plus} \textbf{\bibinfo{volume}{128}},
  \bibinfo{pages}{22} (\bibinfo{year}{2013}).

\bibitem[{\citenamefont{Trachenko}(2017)}]{cycle5} \bibinfo{author}{\bibfnamefont{K.}~\bibnamefont{Trachenko}},
  \bibinfo{journal}{Phys. Rev. D} \textbf{\bibinfo{volume}{95}},
  \bibinfo{pages}{043522} (\bibinfo{year}{2017}).

\bibitem[{\citenamefont{Kragh}(2009)}]{cycle-review} \bibinfo{author}{\bibfnamefont{H.}~\bibnamefont{Kragh}},
  \bibinfo{journal}{Science in Context} \textbf{\bibinfo{volume}{22}},
  \bibinfo{pages}{587} (\bibinfo{year}{2009}).

\bibitem[{\citenamefont{Egan and Lineweaver}(2010)}]{egan-bh} \bibinfo{author}{\bibfnamefont{C.~A.} \bibnamefont{Egan}} \bibnamefont{and}
  \bibinfo{author}{\bibfnamefont{C.~H.} \bibnamefont{Lineweaver}},
  \bibinfo{journal}{The Astrophys. J.} \textbf{\bibinfo{volume}{710}},
  \bibinfo{pages}{1825} (\bibinfo{year}{2010}).

\end{thebibliography}

\end{document}